\documentclass{aastex631}

\usepackage{amsmath}

\begin{document}

\title{Constraining anisotropic diffusion between Geminga and Earth with the cosmic-ray electron and positron spectrum}

\author{Junji Xia}
\affiliation{The School of Physical Science and Technology, Southwest Jiaotong University, Chengdu, 611756, China}

\author{Xiaojun Bi}
\affiliation{Key Laboratory of Particle Astrophysics, Institute of High Energy Physics, Chinese Academy of Sciences, Beijing 100049, China}
\affiliation{University of Chinese Academy of Sciences, Beijing 100049, China}

\author{Kun Fang}
\affiliation{Key Laboratory of Particle Astrophysics, Institute of High Energy Physics, Chinese Academy of Sciences, Beijing 100049, China}

\author{Siming Liu}

\affiliation{The School of Physical Science and Technology, Southwest Jiaotong University, Chengdu, 611756, China}
\begin{abstract}

The gamma-ray halo around Geminga indicates significant suppression of cosmic-ray diffusion. One possible explanation for this phenomenon is the projection effect of slow diffusion perpendicular to the mean magnetic field (characterized by the diffusion coefficient $D_\perp$) within an anisotropic diffusion framework. In this scenario, the diffusion coefficient parallel to the mean field ($D_\parallel$) can still be large, enabling electrons and positrons ($e^\pm$) produced by Geminga to efficiently travel to Earth along the magnetic field lines, possibly resulting in a detectable $e^\pm$ flux. In this work, we first determine the basic parameters of the anisotropic model using the morphology and spectral measurements of the Geminga halo and then predict the flux of $e^\pm$ produced by Geminga at the location of Earth. We find that the $e^-+e^+$ spectrum of DAMPE can give crucial constraint on the anisotropic diffusion model: to ensure that the predicted spectrum does not exceed the measurements, the Alfv\'{e}n Mach number of the turbulent magnetic field ($M_A$) should not be less than $0.75$, corresponding to $D_\parallel/D_\perp\lesssim3$ given that $D_\perp=D_\parallel M_A^4$. This implies that a significant suppression of $D_\parallel$ relative to the average value in the Galaxy may still be necessary. Furthermore, we find that under the anisotropic diffusion model, Geminga can produce a very sharp feature around $1$ TeV in the $e^-+e^+$ spectrum, which could naturally explain the peculiar $1.4$ TeV excess tentatively observed by DAMPE.


\end{abstract}


\section{Introduction} \label{sec:intro}
Pulsars are important factories for producing high-energy electrons and positrons in the universe \citep{1969ApJ...157..869G}. On the surface of a pulsar, seed electrons are extracted from the crust and accelerated by a potent electric field induced by the rotation of the pulsar. These electrons, moving along magnetic field lines, emit photons via curvature radiation, leading to the production of numerous electron-positron pairs when the emitted photons interact with the magnetic field. Subsequently, the electrons and positrons propagate outward as part of a highly relativistic pulsar wind, experiencing further acceleration at the termination shock of the surrounding pulsar wind nebula \citep[PWN]{Rees:1974nr}.

The duration of electron acceleration in a PWN is significantly longer than that in a supernova remnant (SNR). After the initial PWN is disrupted by the reverse shock of the SNR, the pulsar gradually leaves its initial PWN and interacts with the ISM, forming a new bow-shock PWN \citep{Bykov:2017xpo}. Bow-shock PWNe with ages in the order of $100$~kyr can still accelerate electrons to $\sim100$~TeV, such as the Geminga PWN \citep{caraveo2003geminga}. High-energy electrons and positrons are proposed to escape from the bow-shock tail \citep{10.1093/mnras/sty2237} and diffuse into the ISM. In some cases, the escaped electrons and positrons can generate bright gamma-ray halos around the central PWN through inverse Compton (IC) scattering of background photons \citep{PhysRevD.100.043016,liu2022physics,fang2022gamma,lopez2022gamma}. This phenomenon is referred to as pulsar halos. 

An intriguing characteristic revealed by pulsar halos is the slow-diffusion phenomenon of electrons and positrons. Specifically, the diffusion coefficient of electrons and positrons inferred from their surface brightness distribution is two orders of magnitude smaller than the typical value in the Galaxy \citep{abeysekara2017extended,Aharonian:2021jtz,Fang:2022qaf,HAWC:2023jsq}. The slow-diffusion phenomenon may be interpreted by the turbulent environment created by the SNR associated with the pulsar \citep{Kun:2019sks}, the turbulent magnetic fields amplified by streaming instability of escaping electrons and positrons themselves \citep{Evoli:2018aza,Mukhopadhyay:2021dyh}, or the projection effect of diffusion perpendicular to the mean magnetic field direction under the anisotropic diffusion scenario \citep{liu2019understanding,DeLaTorreLuque:2022chz,fang2023effect}.

Regardless of which of the above models is considered, the escaping electrons and positrons will not be confined solely near the pulsar, which means that electrons and positrons produced by nearby pulsars, such as Geminga and Monogem, still have the possibility to generate a significant flux at Earth. In the first two scenarios, slow diffusion is expected to occur only within a few tens of pc from the pulsar, which is also supported by current observations \citep{Fang:2021qon,Fang:2023xla}. There has been considerable discussion on explaining the electron and positron spectra at Earth from nearby pulsars like Geminga under the assumption of such a two-zone diffusion model \citep[e.g.]{Hooper:2017gtd,Fang:2018qco}. For the anisotropic diffusion model, electrons and positrons can escape to distant regions in a direction parallel to the magnetic field. However, its implications for electron and positron observations have not yet been explored.

In this work, we discuss the contribution of Geminga to the cosmic-ray positron spectrum and the total electron-positron ($e^-+e^+$) spectrum under the anisotropic diffusion model. The measurements of these spectra can, in turn, provide important constraints on the anisotropic diffusion model. In Section~\ref{sec:method}, we introduce the calculation methods involved in this study. In Section~\ref{sec:limit}, we first determine some of the model parameters using the HAWC observations of the Geminga halo and then compare the model-predicted positron and $e^-+e^+$ spectra with the AMS-02 and DAMPE measurements to further constrain the model. Furthermore, we find that under the anisotropic diffusion model, Geminga may produce a very sharp spectral feature around $1$~TeV in the $e^-+e^+$ spectrum. In Section~\ref{sec:dampe}, we discuss the possibility that this feature could explain the anomalous excess around $1.4$~TeV observed in the $e^-+e^+$ spectrum by DAMPE. Section~\ref{sec:conclu} is the conclusion of the work.



\section{Method} \label{sec:method}
After electrons and positrons escape from a PWN, their propagation in the ISM can be modeled using the diffusion-loss equation. We first solve the anisotropic propagation equation to determine the number density distribution of electrons and positrons released by the Geminga PWN. Subsequently, we perform the line-of-sight (LOS) integral on the number density and use the standard IC scattering calculation to obtain the gamma-ray surface brightness around Geminga.

The temporal evolution of the particle number density, N, in both spatial and energy space, is governed by the transport equation:
\begin{equation}
\begin{aligned}
\frac{\partial N(E_e, r_\parallel, r_\perp, t)}{\partial t} =  D_\parallel(E_e) \frac{1}{r_\parallel} \frac{\partial}{\partial r_\parallel} \left[ r_\parallel \frac{\partial N(E_e,  r_\parallel, r_\perp, t)}{\partial r_\parallel} \right] + D_\perp(E_e) \frac{\partial^2 N(E_e,  r_\parallel, r_\perp, t)}{\partial r_\perp^2}  \\ + \frac{\partial}{\partial E_e} \left[ b(E_e) N(E_e,  r_\parallel, r_\perp, t) \right] + Q(E_e,  r_\parallel, r_\perp, t),
\end{aligned}
\end{equation}

where $N$ is the electron number density, $E$ is the electron energy, $r_\parallel$ is the coordinate parallel to the mean magnetic field with the pulsar position as the origin, $r_\perp$ is the coordinate perpendicular to the mean magnetic field with the pulsar position as the origin, and $t$ is the time coordinate with the pulsar birth time as the origin. We assume that the mean magnetic field in the interstellar medium between Earth and Geminga is uniform. As suggested by \citet{xu2013cosmic} and \citet{yan2008cosmic}, the diffusion coefficients perpendicular and parallel to the mean magnetic field are related by
\begin{equation} \label{eq:diffusion}
\newcommand{\parallelsum}{\mathbin{\!/\mkern-5mu/\!}}
D_{\perp} =M_A^4 D_{\parallel}= D_{0\perp,}\cdot(\frac{E}{100\;\rm{TeV}})^{1/3}
\end{equation}
where $M_A$ is the Alfvénic Mach number. Q is the source term depicting the electron injection from the pulsar:
\begin{equation}
Q(E,t)=Q_{1}(t)(\frac{E}{\rm{GeV}})^{-\alpha}\exp(-\frac{E}{E_\text{cut}})
\end{equation}
\begin{equation}
Q_{1}(t)=\frac{N_0}{(1+t/T_0)^{n+1/n-1}}
\end{equation}
In this context, $N_0$ denotes the normalization constant, $\alpha$ represents the spectral index, and  $E_{cut}$ corresponds to the high-energy cutoff of the injection spectrum.
The energy-loss rate is denoted by \( b(E) = b_0(E)E^2 \). We take a magnetic field strength of 3 $\mu$G for 
the synchrotron loss rate. In addition to considering the cosmic microwave background(with a temperature of 2.7 K and energy density of 0.25 eV cm$^{-3}$)\citep{abeysekara2017extended}, we adopt an infrared photon field (with a temperature of 20 K and energy density of 0.3 eV cm$^{-3}$), and an optical photon field (with a temperature of 5000 K and energy density 0.3 eV cm$^{-3}$) for the IC energy-loss rate, and the Klein-Nishina effect is also taken into account \citep{fang2021klein}. Then we can derive the electron and positron distribution by a coordinate transformation method \citep{fang2023effect}, which is expressed as:
\begin{equation}
N(E)=\int_{0}^{t_1}dt\frac{c}{4\pi}\frac{Q(E,t)E_{0}(E,t)^{2-\alpha}\exp[\frac{-E_{0}(E,t)}{E_\text{cut}}]}{8\pi^{\frac{3}{2}}E^{2}L_{\bot}^{2}(E,t)L_{\parallel}(E,t)}\exp[-\frac{r_{\parallel}^2}{4L_{\parallel}^{2}(E,t)}-\frac{r_{\bot}^2}{4L_{\bot}^{2}(E,t)})
\end{equation}
Where $E_0$ is the initial energy of an electron with energy $E$ at time $t$, and $t_1$ stands for $1/b_0E$ or the age of the pulsar. 
\begin{equation}
E_{0}\approx \frac{E}{[1 - b_{0} E (t - t_{0})]} \, 
\end{equation}
$L$ is the characteristic distance of diffusion:
\begin{equation}
L(E,t)=\sqrt\frac{D_{0}(E^{\delta-1}-E_{0}^{\delta-1})}{(1-\delta)b_0}
\end{equation}

We denote the angle between the \( z \)-axis and the direction to the pulsar by \( \Phi \). For LOS direction at an angle \( \theta \) relative to the pulsar, the electron surface density can be determined by
\begin{equation}
N_1(E,\theta,\Phi) = \int N(E,r(l, \theta, \Phi), z(l, \theta, \Phi)) \, dl
\end{equation}
For a detailed description of the transformation between the coordinates \((l, \theta, \Phi)\) and \((r, z)\), the reader is referred to \citet{liu2019understanding}. The IC surface brightness can be obtained by
\begin{equation} \label{eq:sbp}
I_{\text{IC}}(E, \theta, \Phi) = \frac{1}{4\pi} \mathcal{F}_{\text{IC}}\{N_1(E,\theta, \Phi), n_{\text{ph}}\}
\end{equation}
respectively, where \( n_{\text{ph}} \) is the photon number density of the background radiation. We adopt the \textit{naima} package \citep{zabalza2015naima} to calculate the IC flux $\mathcal{F}_{\text{IC}}$.

As no significant asymmetry is reported for the Geminga halo in the initial HAWC measurements, the angle between LOS and the direction of the local magnetic field should be small. We assume $\Phi=0$ at the current time. However, due to the significant proper motion of the Geminga pulsar, it has moved about $18^\circ$ from its birthplace on the celestial sphere \citep{faherty2007trigonometric}. This means that in our cylindrical coordinate system, the $r_\perp$-coordinate of Geminga has a significant time evolution. We have accordingly taken into account its transverse velocity in the calculation of the electron number density. Since there are currently no definite constraints on its radial velocity, we assume it to be zero. Thus, in calculating the electron number density, we assume a linear relationship between the age of the pulsar and the angle of incidence. We also calculate the result for \(\phi = 5^\circ\), and find that the profile at this angle is essentially the same as at \(\phi = 0^\circ\).

\section{Constraints on the anisotropic diffusion model} \label{sec:limit}

\begin{deluxetable}{lcc}
\tabletypesize{\scriptsize}
\tablewidth{0pt} 
\tablecaption{Parameters and their values \label{tab:parameters}}
\tablehead{
\colhead{Quantity} & \colhead{Value} & \colhead{Comment}
} 
\startdata 
\multicolumn{3}{c}{\textbf{Time independent quantity}} \\
$E_{\text{cut}}$ & $100$ \text{TeV} & Parameter \\
$\alpha$ & $1$ & Parameter \\
$\delta$ & $1/3$ & Parameter \\
$T_{\text{CMB}}$ & $2.73 \, \text{K}$ & Observed \\
$u_{\text{CMB}}$ & $0.25 \, \text{eV} \, \text{cm}^{-3}$ & Observed \\
$T_{\text{FIR}}$ & $20 \, \text{K}$ & Parameter \\
$u_{\text{FIR}}$ & $0.3 \, \text{eV} \, \text{cm}^{-3}$ & Parameter \\
$T_{\text{OPT}}$ & $5000 \, \text{K}$ & Parameter \\
$u_{\text{OPT}}$ & $0.3 \, \text{eV} \, \text{cm}^{-3}$ & Parameter \\
$D_{\text{0}}$ & $3.2\times10^{27} \, \text{} \, \text{cm}^{-2}{s}^{-1}$ & Parameter \\
$T_{\text{conv}}$ & $ 12 \, \text{kyr}$ & Parameter \\
$L_{\text{now}}$ & $3.26 \times 10^{34} \, \text{erg} \, \text{s}^{-1}$ & Observed \\
$B_{\text{compact}}$ & $3 \, \mu\text{G}$ & Parameter \\
$n$ & $3$ & Assumed \\
$T_{\text{age}}$ & $342 \, \text{kyr}$ & Parameter \\
\enddata
\end{deluxetable}
\begin{figure}
{\includegraphics[scale=0.3]{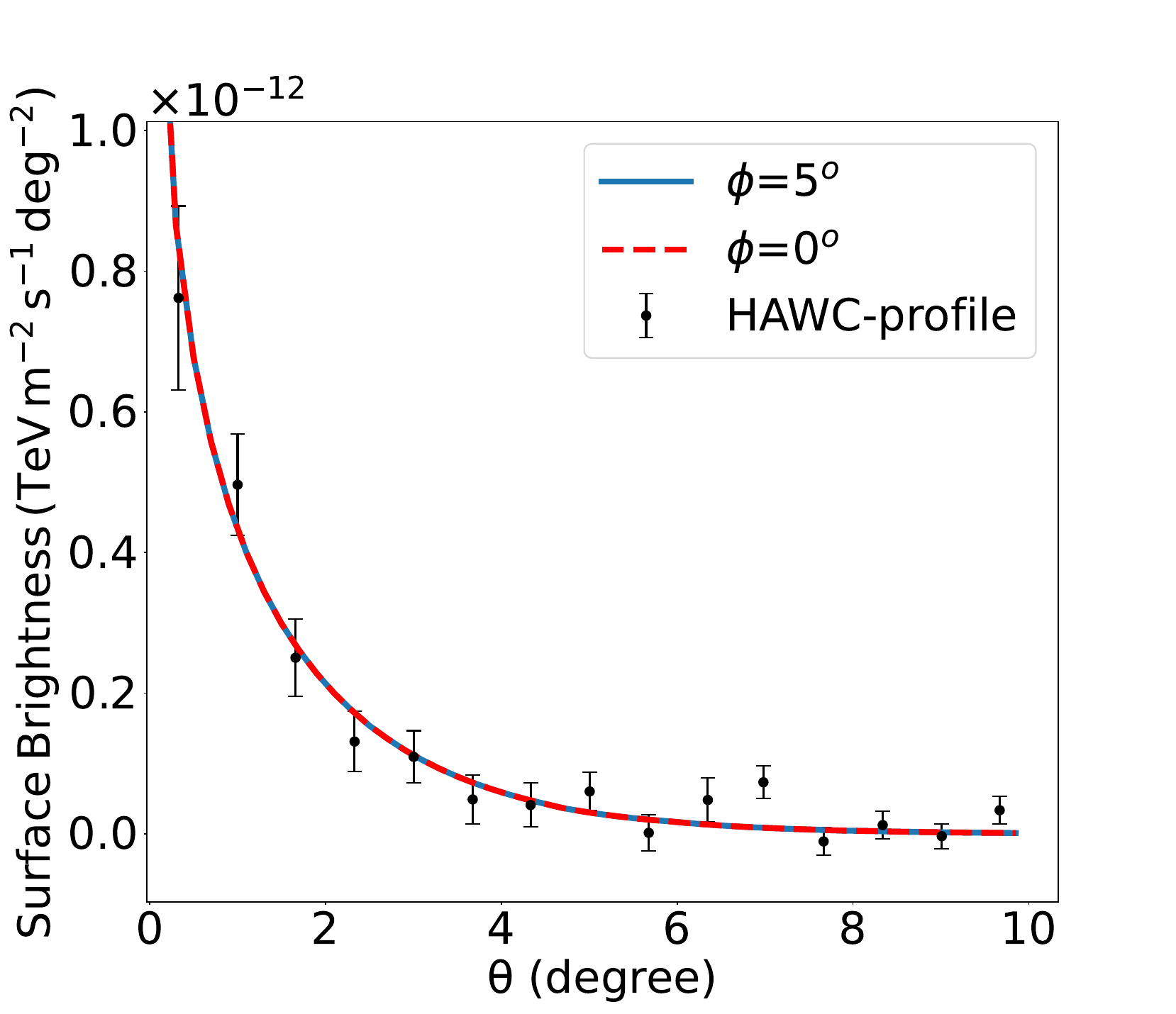}}
{\includegraphics[scale=0.3]{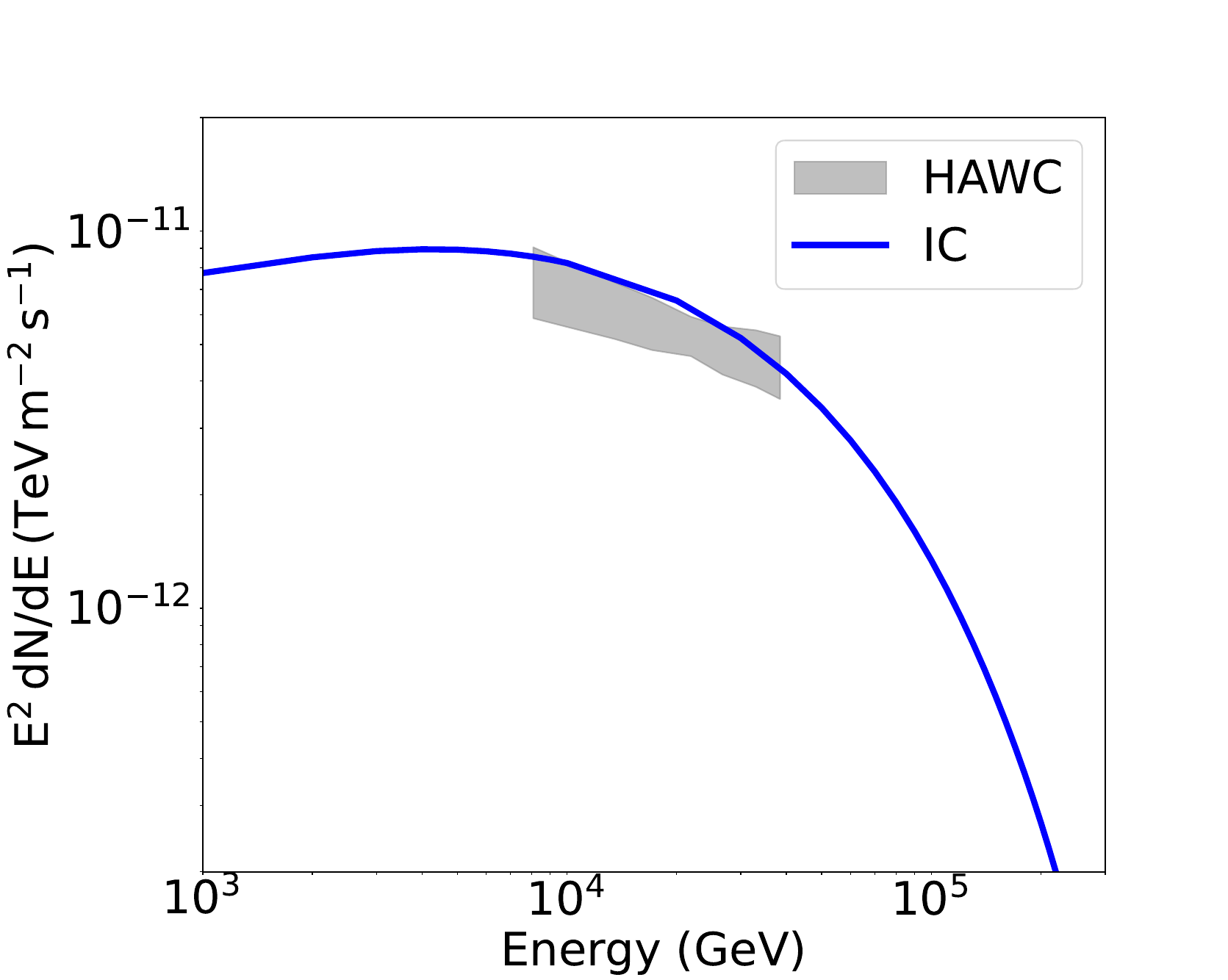}}
	\caption{Left: fitting result for the Geminga halo profile measured by HAWC \citep{abeysekara2017extended} with $\eta$= 0.04 and $M_A$=0.75 (we will show in Section~\ref{sec:limit} that such a large $M_A$ is favored by the measurements of $e^-+e^+$ spectrum). Right: fitting result of the HAWC measurement of the Geminga energy spectrum, with $\eta=0.04$ and $M_A=0.75$.}
	\label{FIG:profile}
\end{figure}
\begin{figure}
\includegraphics[scale=0.35]{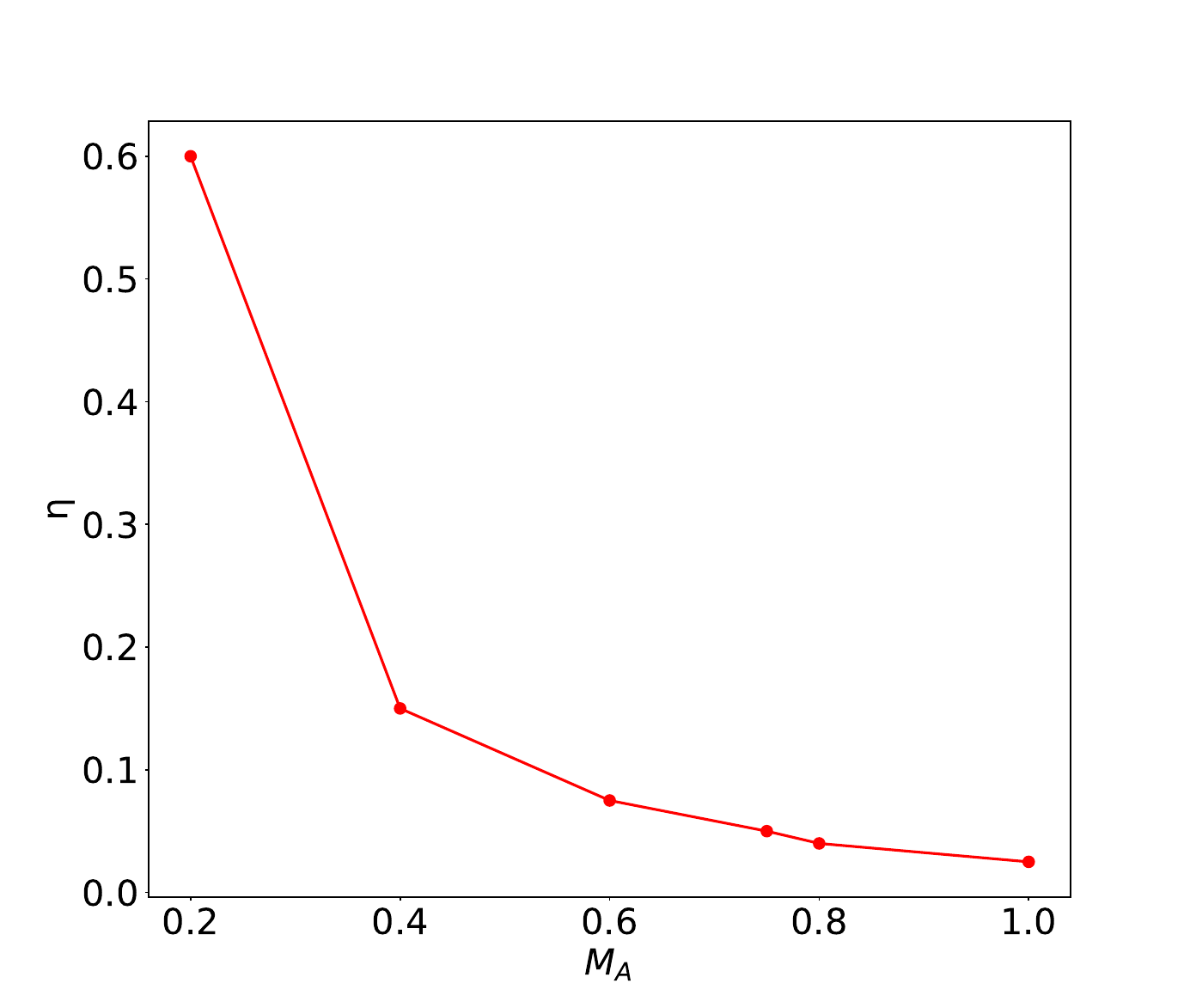}
\includegraphics[scale=0.29]{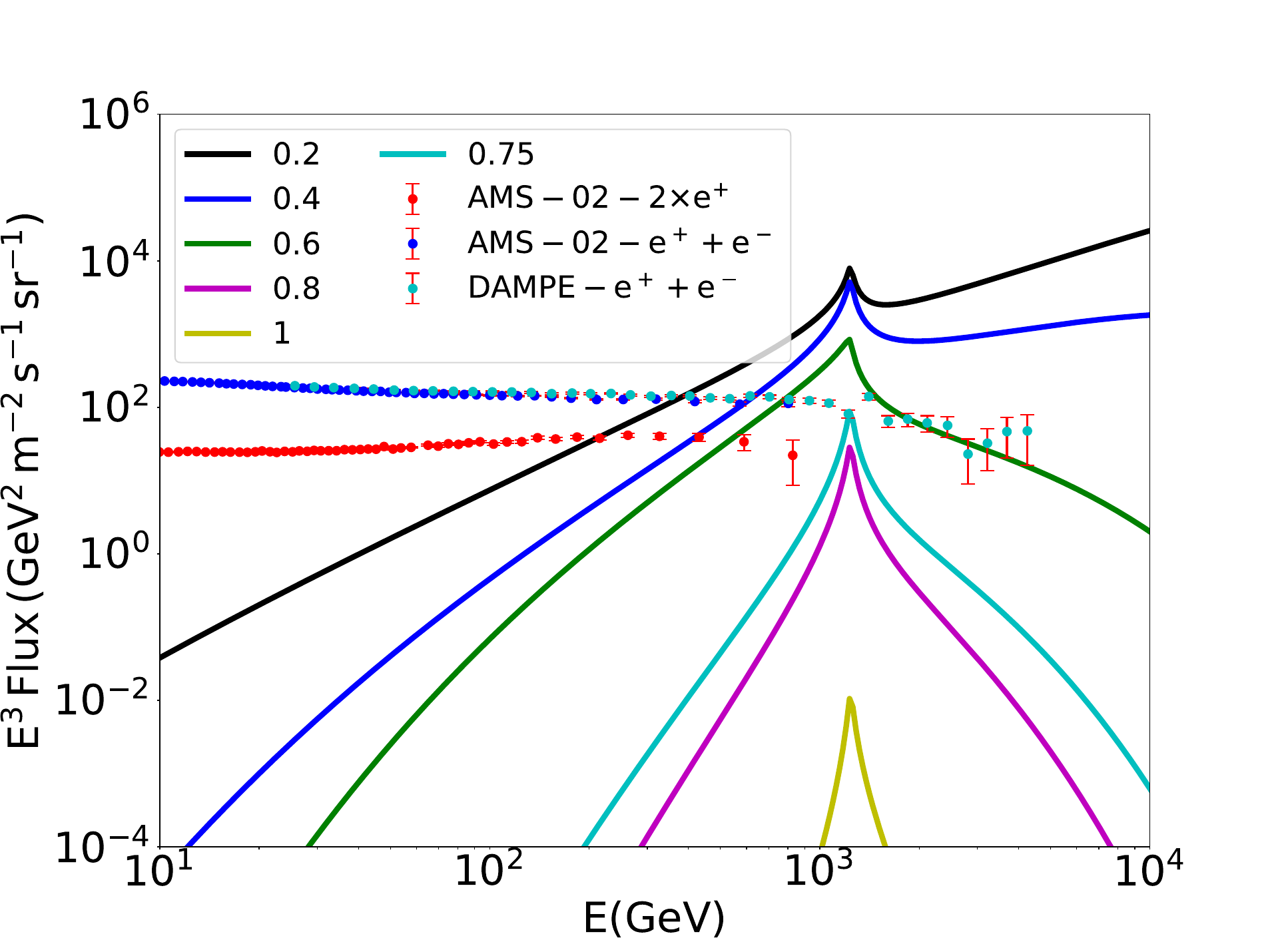}
    \caption{Left: Through the fitting of the Geminga profile and flux observation data, a discernible correlation between the conversion parameter $\eta$ and the Alfv\'{e}nic Mach number $M_A$ is established. Right: The black data points represent the DAMPE $e^-+e^+$ observation data, the blue data points represent the AMS-02 $e^-+e^+$ observation data, and the red data points represent the AMS-02 positron observation data. The different colored lines indicate the electron flux intensity generated under the fitted energy spectrum and profile of the Geminga halo.}
\label{FIG:zixing}
\end{figure}

Based on equation~(\ref{eq:sbp}), we calculate the one-dimensional surface brightness profile of the Geminga halo and the gamma-ray spectrum for $\theta<10^\circ$. By comparing these with the HAWC measurements, we determine the main parameters of the model. The comparison results are shown in Figure~\ref{FIG:profile}, and the parameters used in the calculations are listed in Table~\ref{tab:parameters}. A cutoff energy of $E_{\rm cut}$ is assumed for the electron injection spectrum \citep{bao2022slow}, which is suggested by the updated spectrum measurements of HAWC \citep{2019ICRC...36..832Z}. The perpendicular diffusion coefficient $D_\perp$ is assumed to be the value measured by \citet{abeysekara2017extended}, that is, $3.2 \times10^{27}$ cm$^2$ s$^{-1}$ at 100 TeV. 

When interpreting the profile and energy spectrum of the halo, there is a degeneracy between the Alfv\'{e}n Mach number ($M_A$) and the conversion efficiency from the pulsar spin-down energy to the electron energy ($\eta$). Since $D_\perp$ is constrained by profile measurements, a smaller $M_A$ implies a larger $D_\parallel$ (see equation (\ref{eq:diffusion})), resulting in a higher proportion of electrons propagating beyond the line of sight. Consequently, to maintain the same gamma-ray surface brightness, a larger total number of injected electrons is required, which corresponds to a higher $\eta$. The left panel of Figure \ref{FIG:zixing} illustrates this trend.

The right panel of Figure \ref{FIG:zixing} shows the $e^-+e^+$ spectrum from Geminga reaching Earth under different combinations of $M_A$ and $\eta$ obtained above. The initial work by HAWC has already indicated that electrons and positrons diffuse too slowly to reach Earth in sufficient amounts under the assumption of isotropic diffusion \citep{abeysekara2017extended}. However, in the case of anisotropic diffusion, electrons and positrons can rapidly travel through channels parallel to the magnetic field, potentially making a significant contribution to the observed electron and positron fluxes. As can be seen from the figure, the smaller the $M_A$, meaning larger $D_\parallel$, the higher the expected electron and positron fluxes at Earth.

In the following, we provide a qualitative analysis of the features in the right panel of Figure \ref{FIG:zixing}. The electron and positron spectra calculated with different parameters all exhibit a peak around $1$ TeV. One reason for this is the energy loss of electrons injected at the early stages of the pulsar \citep{Yuksel:2008rf}. The lifetime of electrons with energy $E$ is approximately $t_{\rm cool}=1/(b_0E)$. When $t_{\rm cool}$ is shorter than the pulsar age $T_{\rm age}$, the electrons injected at the earliest stages with energies greater than $E_{\rm crit}\equiv1/(b_0T_{\rm age})$ have significantly diminished at the current time, where $E_{\rm crit}$ is around $1$~TeV. As a result, the electron spectrum above $E_{\rm crit}$ is dominated by electrons injected at later times, leading to a significant drop in electron flux above $E_{\rm crit}$. Another crucial factor determining the spectral features is the characteristic propagation distance of electrons, $L_\parallel$, which can be approximated as $L_\parallel\sim\sqrt{D_\parallel t_E}$, where $t_E={\rm min}\{T_{\rm age}, t_{\rm cool}\}$. When $T_{\rm age}<t_{\rm cool}$, $L_\parallel$ is approximately proportional to $\sqrt{E^{\delta/2}}$, otherwise $\sqrt{E^{(\delta-1)/2}}$. Consequently, in the case of $\delta<1$, $L_\parallel$ reaches its maximum at $T_{\rm age}=t_{\rm cool}$, i.e., $E=E_{\rm crit}$. When $L_\parallel \lesssim r_\parallel$, the exponential term in equation (\ref{eq:diffusion}) exerts a significant suppression effect. Consequently, the electron spectrum experiences a significant decline on both sides of $E_{\rm crit}$, which is another important reason for the peak around $1$~TeV. When $L_\parallel$ increases along with $M_A$, the exponential suppression in equation (\ref{eq:diffusion}) becomes less significant. As shown by the black line in the right panel of Figure \ref{FIG:zixing}, the spectrum on both sides of the peak extends in a power-law manner, and the peak primarily results from the energy-loss effect of electrons mentioned above.

In the right panel of Figure \ref{FIG:zixing}, we compare the $e^-+e^+$ spectra measured by AMS-02 and DAMPE, as well as twice the AMS-02 positron spectrum\footnote{We assume that Geminga produces equal amounts of electrons and positrons, so we can use twice the positron spectrum to constrain the $e^-+e^+$ spectrum produced by Geminga.}, with the model predictions. These measurements can be considered as upper limits on the $e^-+e^+$ spectra produced by the current model, thereby constraining the parameters of the anisotropic diffusion model. Since the total $e^-+e^+$ spectrum is predominantly contributed by electrons from SNR, the AMS-02 positron spectrum theoretically should impose stronger constraints on the model. However, due to the predicted spectrum being very sharp around 1 TeV and the current positron spectrum measurements only extending up to $\approx800$ GeV, it is actually the total $e^-+e^+$ spectrum that imposes stricter constraints on the model. It can be seen that when $M_A\geq0.75$, the model-predicted positron flux does not exceed the measurements. Under this constraint, the difference between $D_\parallel$ and $D_\perp$ is less than a factor of $3$, indicating that the diffusion coefficient in the direction parallel to the magnetic field still needs to be significantly smaller than the average value in the Galaxy. If future AMS-02 measurements can extend the positron spectrum beyond $1$ TeV, it could impose even stricter constraints on the model.

\section{Reproducing the 1.4 TeV spike of DAMPE} \label{sec:dampe}

Intriguingly, in the case where $M_A\geq0.75$, the anisotropic diffusion model predicts a spike feature around $1$ TeV, which brings to mind the sharp spectral bump observed by DAMPE at $1.4$ TeV. Although this spectral feature is still tentative, numerous studies have already discussed its possible origins (e.g., \citet{Yuan:2017ysv,Sui:2017qra,Zhao:2019jjx,Bao:2020ila}). In this section, we discuss the potential explanation for this peculiar spectral structure from Geminga within the framework of the anisotropic diffusion model.

\begin{figure}
{\includegraphics[scale=0.30]{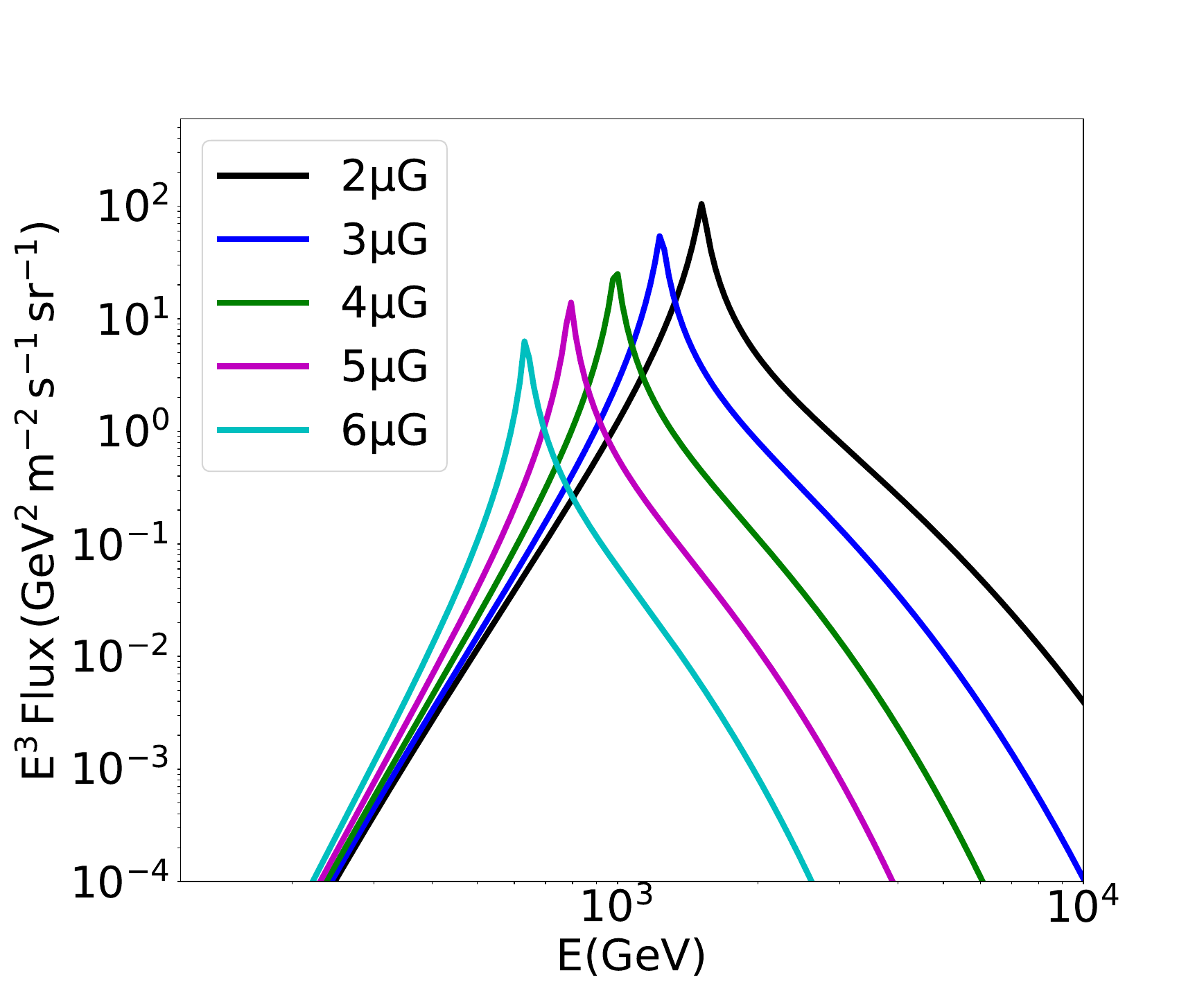}}
{\includegraphics[scale=0.31]{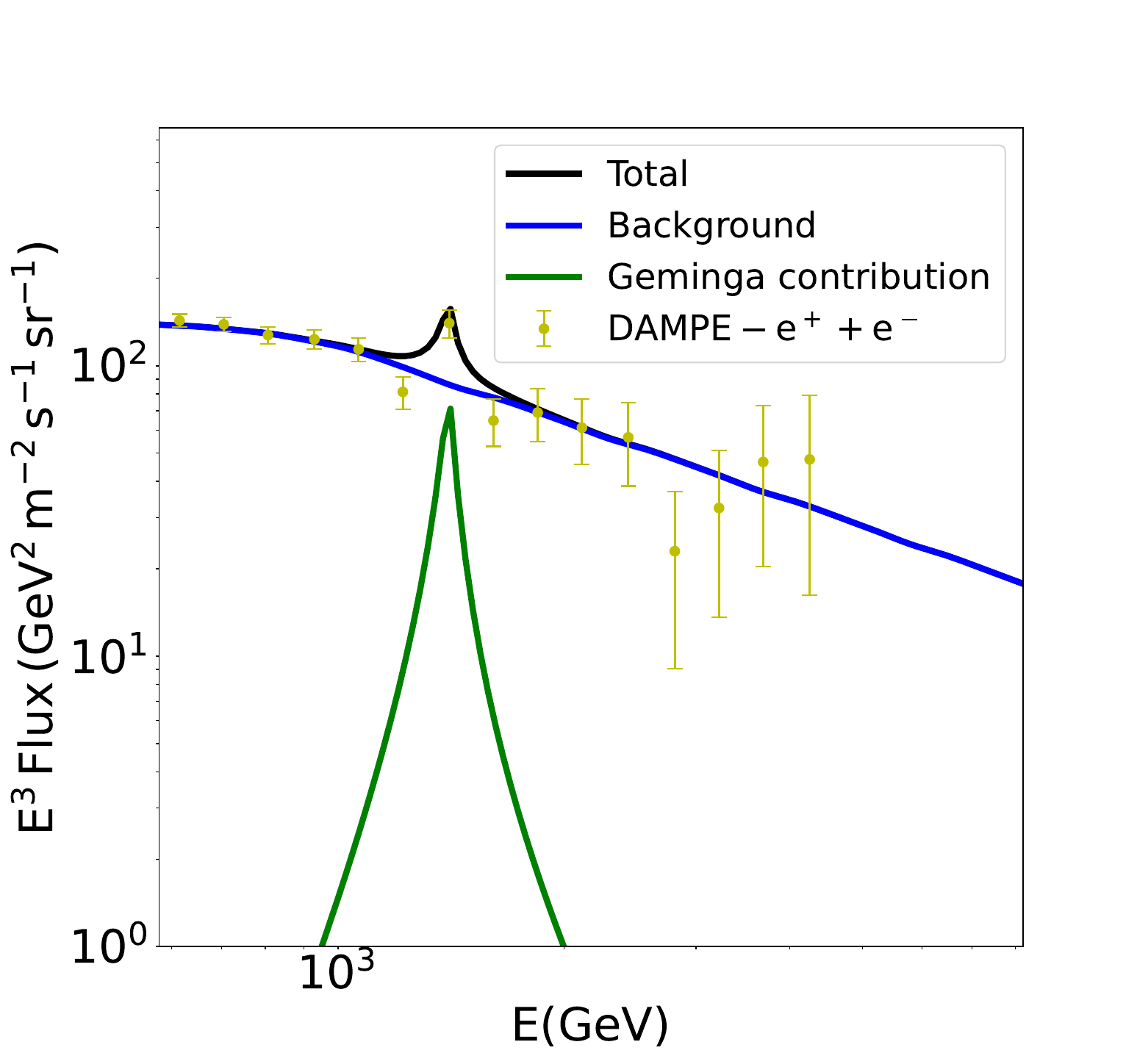}}
\caption{Left: $e^-+e^+$ spectra assuming different magnetic fields. Right: A predicted $e^-+e^+$ spectrum that can accommodate the DAMPE measurement, where $M_A = 0.8$, $\eta = 0.03$, and $B=2.4$ $\mu$G. This set of parameters can also explain the HAWC observations of the Geminga halo. The background electrons are contributed by other sources such as SNRs.}
\label{FIG:DAMP}
\end{figure}

We superimpose the current model on a smooth background component to reproduce the $e^-+e^+$ spectrum measured by DAMPE. For the background component, we adopt a 1D diffusion model to describe cosmic-ray transport along Galactic magnetic field (GMF) lines, as discussed by \citet{schwadron2014global}. This 1D model remains applicable for diffusion within the Galactic disk, provided the radial gradient of CRs is significantly less pronounced than the vertical gradient. The steady-state transport equation for the cosmic-ray distribution function $N(z, E)$ along the GMF is formulated as follows:
\begin{equation}
D(E) \frac{\partial^2 N}{\partial z^2} + \frac{\partial [b(E) N]}{\partial E} + Q(z, E) = 0,
\end{equation}
where the diffusion coefficient is $D(E) = D_1 (E/1 \, \text{GeV})^\delta$, and $D_1$ takes the mean diffusion coefficient of the Galaxy $3 \times10^{28}$ cm$^2$ s$^{-1}$. For cosmic-ray electrons, energy losses are dominated by synchrotron and inverse Compton radiations with the energy loss rate $\dot{E} \equiv -b(E) = -b_0 E^2$.
The Galaxy is conceptualized as a disc with a half-thickness $h$ and a halo with a half-thickness $H > h$, oriented along a vertical magnetic field. Neglecting the diffusion perpendicular to the magnetic field renders the radial structure of the disc inconsequential. A homogeneous source term $Q(z, E) = q(E) \theta(h - |z|)$ is considered, where $\theta(x)$ denotes the Heaviside step function and $q(E)$ represents the injection spectrum. It is assumed that there are no cosmic rays at the halo boundary, such that $N(z = \pm H, E) = 0$.
One can solve equation (9) via Fourier series \citep{shi2019origin}:
\begin{equation}
\begin{aligned}
N(z, E) = & \frac{2}{b_0 E^2} \sum_{m=0}^{\infty} \sin \left[ \frac{\left( m + \frac{1}{2} \right) \pi H}{(m + \frac{1}{2}) \pi} \right] \cos \left[ \left( m + \frac{1}{2} \right) \frac{\pi z}{H} \right] \\
& \times \int_{E}^{\infty} dE' q(E') \exp \left\{ \left[ \left( m + \frac{1}{2} \right) \frac{\pi l(E)}{H} \right]^2 \left[ \left( \frac{E}{E'} \right)^{1-\delta} - 1 \right] \right\}.
\end{aligned}
\end{equation}
At a given energy $E$, there is characteristic length
\begin{equation}
\ell(E) \equiv \sqrt{\frac{D_0 E^{\delta - 1}}{(1 - \delta) b_0}},
\end{equation}

For the $e^-+e^+$ spectrum produced by Geminga, we have explained in the previous section that it will peak at $E_{\rm crit}$. The value of $E_{\rm crit}$ is influenced by the energy loss factor $b_0$, which can be affected by the magnetic field strength $B$ in the interstellar medium. In the left panel of Figure \ref{FIG:DAMP}, we show the variation of the spectral peak produced by Geminga with changes in $B$. As the magnetic field strength increases, the position of the peak shifts to lower energies. Due to the uncertainty in the magnetic field strength in the nearby interstellar medium, the expected peak energy may also vary within a certain range.

As shown in the right panel of Figure \ref{FIG:DAMP}, when the magnetic field is $2.4$ $\mu$G, the spectrum produced by Geminga, when superimposed on a smooth background component, can well reproduce the spike observed by DAMPE at $1.4$ TeV, both for the normalization and position of the spike. In this case, the assumed values are $M_A = 0.8$ and $\eta = 0.03$, with the other parameters consistent with those in Table \ref{tab:parameters}. We highlight that this set of parameters is obtained under the constraints of the Geminga halo profile and spectral measurements. This implies that with the current set of parameters, the anisotropic diffusion model can provide a self-consistent interpretation for both the Geminga halo and the spectral spike observed by DAMPE.

\section{CONCLUSION} \label{sec:conclu}
In this study, we investigate the contribution of Geminga to the electron and positron flux at Earth under the anisotropic diffusion scenario and impose constraints on the anisotropic model by comparing it with the electron and positron spectra measured by DAMPE and AMS-02. Since anisotropic diffusion is a possible explanation for the morphology of the Geminga halo, we first constrain the model parameters using the spectral and morphological observations of the Geminga halo. We find that there is a degeneracy between the conversion efficiency from the pulsar spin-down energy to the electron-positron energy ($\eta$) and the Alfv\'{e}n Mach number ($M_A$) of the turbulent magnetic field, indicating that current observations of the Geminga halo cannot simultaneously determine both. Based on the degeneracy relationship between $M_A$ and $\eta$, we predict the $e^-+e^+$ spectrum produced by Geminga at Earth. The results indicate that when $M_A<0.75$, the predicted $e^-+e^+$ spectrum exceeds the current measurements, which favors a relatively large $M_A$.

The electron and positron flux reaching Earth is primarily influenced by the diffusion coefficient along the magnetic field direction, $D_\parallel$. Given the constraint on $D_\perp$ from gamma-ray observations of the Geminga halo, a smaller $M_A$ implies a larger $D_\parallel$ (since $D_\perp=M_A^4D_\parallel$), resulting in an excessive number of electrons and positrons reaching Earth. Under the constraint of $M_A\geq0.75$, the difference between $D_\parallel$ and $D_\perp$ is less than a factor of 3, indicating that $D_\parallel$ still needs to be significantly smaller than the average cosmic-ray diffusion coefficient in the Galaxy.

Furthermore, we find that under the anisotropic diffusion model, Geminga can produce a very sharp feature in the $e^-+e^+$ spectrum around 1 TeV, which is the result of the synergy of energy losses of the early injected electrons and the energy-dependent diffusion distance of electrons. This can naturally reproduce the tentative spike observed by DAMPE in the $e^-+e^+$ spectrum at 1.4 TeV. We show that when $M_A=0.8$, $\eta=0.03$, and $B=2.4$ $\mu$G, the current model can provide a self-consistent interpretation for both the Geminga halo and the spectral spike of DAMPE.

In this work, we have assumed that the magnetic field in the interstellar medium between Earth and Geminga is anisotropic and uniform. We are aware that considering the typical magnetic field correlation length ($\sim100$ pc), the direction of the magnetic field may change significantly over this scale, which may lead to differences between the actual results and our current calculations. Additionally, taking into account the finite correlation length of the magnetic field, the expected morphology of the Geminga halo would be more complex \citep{fang2023effect,Bao:2024rrg}. Future measurements of the Geminga halo by LHAASO may provide constraints on the magnetic field correlation length under the anisotropic diffusion hypothesis, thereby offering more insights into the propagation of electrons and positrons emitted by Geminga.

\section{Acknowledgments}
We would like to acknowledge Dr. En-Sheng Chen and Dr. Peng-Fei Yin for helpful comments. This work is supported by National Natural Science Foundation of China under grant No. 12105292, No. 12175248, No. 12375103, and No. 12393853.


\label{sec:cite}

\bibliography{sample631.bib}{}
\bibliographystyle{aasjournal}



\end{document}